\documentclass[journal=nl,manuscript=article]{achemso}
\usepackage{amsmath}
\usepackage{amssymb}
\usepackage{graphicx}
\usepackage{epsfig}
\usepackage{dcolumn}
\usepackage{bm}
\usepackage{rotating}
\usepackage{multirow}
\usepackage{xspace}
\usepackage[table]{xcolor}
\usepackage[english]{babel}
\usepackage{colortbl}

\def\bk{{\bf k}}
\def\bq{{\bf q}}
\def\bG{{\bf G}}
\def\bQ{{\bf Q}}

\def\krm{\rm K}
\def\kprime{\rm K'}

\author{Yiming Pan}
\affiliation{Institut f\"ur Theoretische Physik und Astrophysik, Christian-Albrechts-Universit\"at zu Kiel, Kiel, Germany} 
\author{Fabio Caruso}
\email{caruso@physik.uni-kiel.de}
\affiliation{Institut f\"ur Theoretische Physik und Astrophysik, Christian-Albrechts-Universit\"at zu Kiel, Kiel, Germany} 
\title{Vibrational dichroism of chiral valley phonons} 

\begin{document}	
\begin{abstract}
Valley degrees of freedom in transition-metal dichalcogenides influence
thoroughly electron-phonon coupling and its nonequilibrium dynamics. We
conducted a first-principles study of the quantum kinetics of chiral phonons
following valley-selective carrier excitation with circularly-polarized light.
Our numerical investigations treat the ultrafast dynamics of electrons and
phonons on equal footing within a parameter-free ab-initio framework. We report
the emergence of valley-polarized phonon populations  in monolayer MoS$_2$ that
can be selectively excited at either the K or $\rm K'$ valleys depending on the
light helicity. The resulting vibrational state is characterized
by a distinctive chirality, which lifts time-reversal symmetry of the lattice
on transient timescales. We show that chiral valley phonons can further lead to
fingerprints of vibrational dichroism detectable by ultrafast diffuse
scattering and persisting beyond 10~ps. The valley
polarization of nonequilibrium phonon populations could be exploited as
information carrier, thereby extending the paradigm of valleytronics to the
domain of vibrational excitations. 
\end{abstract}

\maketitle

\begin{figure*}[t]
\begin{center}
\includegraphics[width=0.6\textwidth]{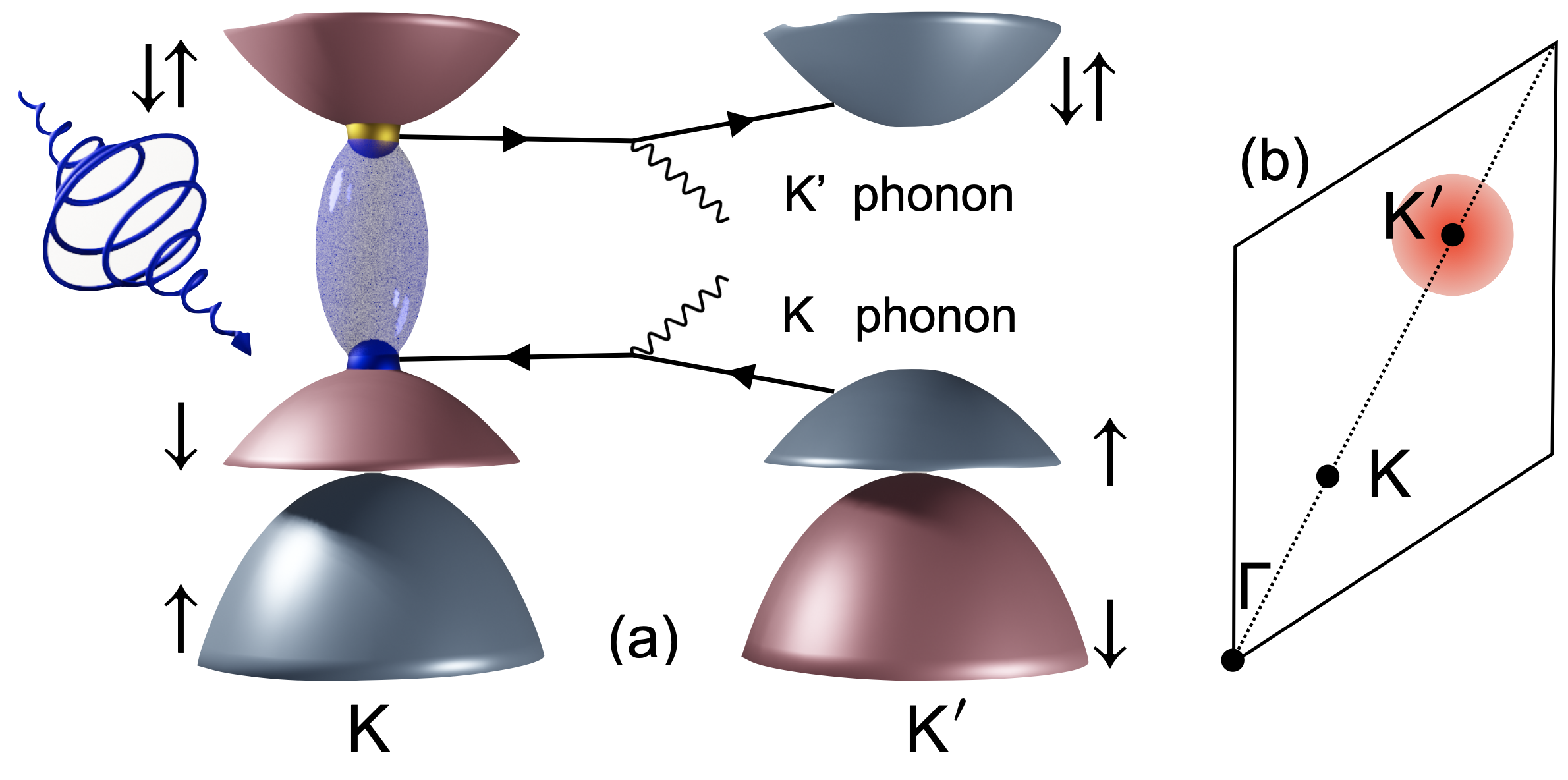}	
\end{center}
\caption{\label{valley} (a) Schematic illustration of the twofold degenerate
valence and conduction bands and of a valley-selective optical excitation
formed upon absorption of circularly-polarized light in monolayer MoS$_2$.
Relaxation of valley-polarized electrons is governed by  electron-phonon
scattering, resulting in the emissions of phonons at K or K$'$. (b) Brillouin
zone and high symmetry points. The red circle denotes the expected momentum
distribution of emitted phonons throughout the thermalization of a
valley-polarized electron distribution.} 
\end{figure*}

The valleytronics paradigm relies on the idea of exploiting valleys degrees of
freedom -- degenerate band extrema in direct band-gap semiconductors -- as a
route to store and manipulate information.\cite{MoS2_valleytronic,Xiao2007}
Two-dimensional transition metal dichalcogenides (TMDs) have heralded
applications in this field, owing to the possibility to selectively control the
carrier population of their twofold degenerate valleys with
circularly-polarized light, and therewith realize a well-defined
valley-polarized state.  In particular, the lack of an inversion center in the
lattice symmetry induces valley-dependent optical selection rules, where
interband optical transitions in the vicinity of the K (K$'$) high-symmetry
point in the hexagonal Brillouin zone are activated by left (right-) handed
circularly polarized
light,\cite{mak2012control,cao2012valley,mak2018light,Jones2013,zeng2012valley,valley_exciton}
as sketched in Fig.~\ref{valley}~(a). Additionally, valley degrees of freedom
are highly sensitive to external stimuli -- as exemplified by the emergence of
Stark,\cite{ye2017optical,Sie2015-hh,Kim2014}
Zeeman,\cite{Aivazian2015,Li2014,MacNeill2015} and Hall
effects,\cite{Mak2014the_valley,Onga2017} -- thus providing numerous
opportunities to realize properties on demand.

Besides electronic properties, valley degrees of freedom further exert a
striking influence on the dynamics of the crystalline lattice. Intervalley
scattering of electrons and holes is a phonon-assisted process accompanied by
the absorption or emission of chiral phonons, i.e., vibrational modes of the
lattice characterized by circular motion of the metal or chalcogen atom with a
definite
helicity.\cite{chiral_phonon_experiment,chiral_phonon_theory,Li2019momenrum}
Chiral phonons are located at either the K or K$'$ points, which confers them a
well-defined valley polarization index. Additionally, they carry angular
momentum and Berry curvature, which are expected to give rise to nontrivial
topological properties, as e.g. the phonon valley Hall
effect,\cite{phonon_hall} or induce novel coupling mechanisms between magnetic
and vibrational degrees of freedom. \cite{Juraschek2022,Xiong2022} Differently
from electrons, chiral valley phonons cannot be excited by photon absorption
due to momentum conservation,\cite{chiral_phonon_experiment} but they can only
emerge as a byproduct of phonon-assisted electron scattering, making them an
inherently nonequilibrium phenomenon.

Pump-probe experiments offer a viable approach to explore the nonequilibrium
dynamics of TMDs following valley-selective photoexcitation, and shed light on
the ultrafast dynamics of chiral phonons.  Experimental studies have thus far
been based on time-resolved Kerr
rotation,\cite{mos2_valley_lifetime,plechinger2014timeresolved,Zhu2014,Dey2017,MoSe2_NEGF,WSe2_NEGF}
photoluminescence,\cite{Lagarde2014,Wang2014,Yan2015}, differential
transmission spectra,\cite{mai2014many,Mai2014,Schmidt2016}  and time- and
angle-resolved photoemission spectroscopy \cite{ws2_valley_lifetime,Kunin2023}
using circularly-polarized pump pulses.  These works demonstrated the
possibility to form electronic excitations with well-defined valley
polarization, shedding light on the timescales and decay channels underpinning
the depolarization dynamics.  The lattice dynamics established throughout the
valley depolarization of electrons and holes still remains uncharted ground for
both experiments and theory. In particular, experimental evidence of the chiral
phonons has been gained only indirectly from the change of light helicity in
infrared spectroscopy.\cite{chiral_phonon_experiment}

First-principles investigations of the valley depolarization dynamics in TMDs
focused primarily on electronic degrees of freedom,  yielding timescales in
excellent agreement with pump-probe experiments and revealing a dominant
contribution of phonon-assisted processes to the
depolarization.\cite{MX2_eph_soc,Xu2021,Lin2022,WSe2_NEGF} The lattice dynamics
however is typically treated at the level of a thermal bath with constant
temperature.  The major challenge pertaining the theoretical modelling of
phonons out of equilibrium is the requirement to account for the coupled
ultrafast dynamics of electrons and phonons in presence of electron-phonon and
phonon-phonon interactions within a first-principles framework, a quest that
only recently has become accessible owing to the progress in the development of
ab-initio codes packages.\cite{Quantum_epsresso,EPW,Pizzi2020,shengBTE} 
        
In this manuscript, we present a theoretical investigation of the ultrafast
phonon dynamics of monolayer MoS$_2$ following absorption of
circularly-polarized light.  Our calculations are based on the time-dependence
Boltzmann equation (TDBE) -- the state of the art for the ultrafast dynamics of
coupled electron-phonon systems -- where the dynamics of electrons and phonons
is treated on an equal footing and without resorting to empirical
parameters.\cite{TDBE,TDBE2,TDBE3,TDBE4,MoS2_TDBE,Jhalani2017} We report the
emergence of  valley-polarized phonon populations and vibrational circular
dichroism in the response of the system, which arises throughout the
phonon-assisted relaxation of valley-polarized electrons and holes.
Valley-polarized phonons can persist beyond 10~ps and reveal the possibility to
control the nonequilibrium dynamics of the lattice via circularly-polarized
light. This novel facet in the lattice dynamics could offer unprecedented
opportunity to exploit valley degrees of freedom in TMDs, and therewith extend
the paradigm of valleytronics to vibrational excitations.  We further predict
the emergence of distinctive fingerprints of vibration dichroism in ultrafast
electron/x-ray scattering experiments, which we expect to foster and facilitate
future experimental studies in this direction.   
        
\begin{figure*}[t]
\begin{center}
\includegraphics[width=0.98\textwidth]{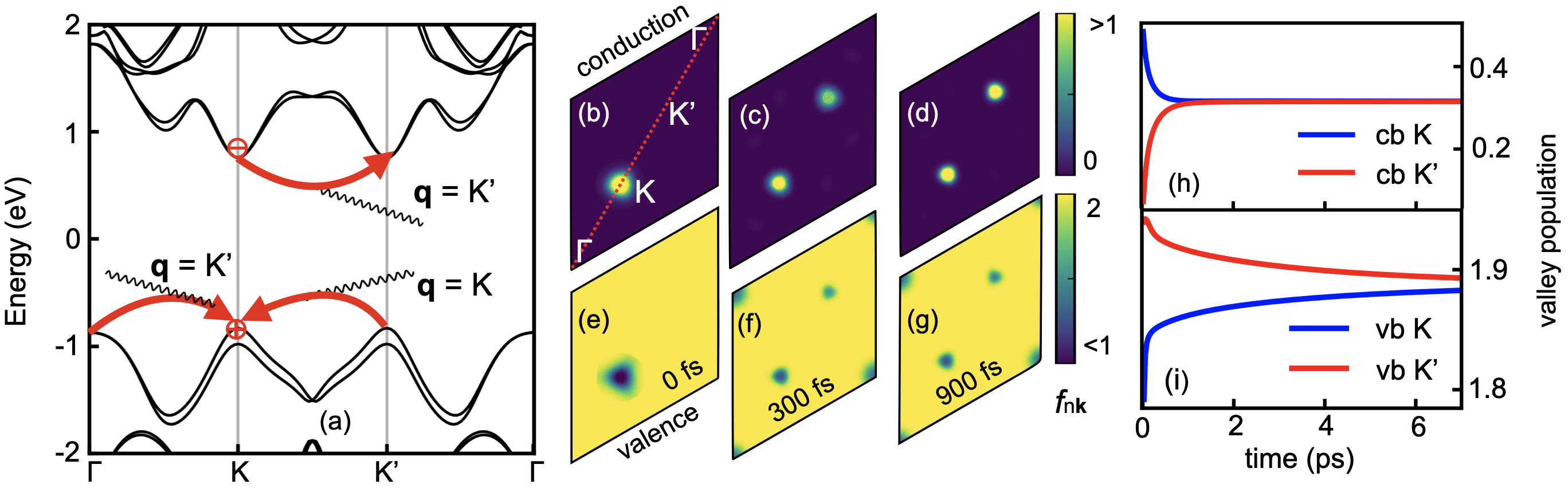}	
\end{center}
\caption{\label{electron_dynamics} Valley depolarization dynamics of electrons
and holes. (a) DFT band structure and intervalley electronic scattering for
valley-polarized electrons and holes. (b)-(d) Time- and momentum-resolved
electron distribution functions $f_{n{\bf k}}(t)$ in the conduction band for
$t=0$, 300, and 900 fs, respectively, for momenta in the Brillouin zone.
(e)-(g) Electron distribution functions $f_{n{\bf k}}(t)$ in the valence band
for the same time snapshots as in panels (b)-(d). (h) Time dependent valley
population at the K and K$'$ high-symmetry points in the conduction bands. (i)
Time dependent valley population of the K and K$'$ high-symmetry points in the
valence band.}
\end{figure*}

The band structure of monolayer MoS$_2$ exhibits a direct band gap and twofold
degenerate valence and conduction bands (valleys) located at the $\krm$ and
$\kprime$ high-symmetry points. A schematic illustration of the band edges is
reported in Fig.~\ref{valley}~(a), whereas the ab-initio band structure, as
obtained from density-functional theory (DFT), is shown in
Fig.~\ref{electron_dynamics}~(a).  Absorption of left-handed (right-handed)
circularly-polarized light results in  a valley-selective optical excitation,
characterized by excited electrons and holes at the K (K$'$) high-symmetry
point.\cite{Yao2008} In the following, we proceed to investigate the ultrafast
dynamics of electrons and phonons triggered by the relaxation of a
valley-polarized electronic excitation. 
    
As initial condition, we consider a valley-polarized state characterized by
photo-excited electrons and holes at the $\krm$ valley in the conduction and
valence band, respectively, whereas carriers in the K$'$ valley are initially
at equilibrium. We assume a photoexcited carrier density of $n=\rm 2\times
10^{13} \ cm^{-2}$.  This choice (schematically depicted in
Fig.~\ref{valley}~(a)) is consistent with the conditions realized in pump-probe
experiments using circularly polarized
light,\cite{mak2012control,zeng2012valley} and with earlier theoretical studies
of the valley depolarization dynamics.\cite{MoSe2_NEGF,WSe2_NEGF} 
The electronic occupations in the K valley are determined by a Fermi-Dirac function
$f_{n\bk}(t=0) = [e^{(\varepsilon_{n\bk}-\mu) /k_BT_e}+1]^{-1}$ with electronic
temperature $T_{\rm e} = 1000$~K. The chemical potential $\mu$ for electrons
and holes is fixed to yield the photoexcited carrier density $n$.  The initial
carrier distribution $f_{n{\bf k}}(t=0)$ in the conduction (valence) bands is
illustrated in Fig.~\ref{electron_dynamics}(b)
[Fig.~\ref{electron_dynamics}(e)]  for crystal momenta in the Brillouin zone.
{\color{black} These initial conditions correspond to an increase of electronic
energy of 41 meV per unit cell which --   
considering an absorption rate of about 10\% -- 
can be established by a pump pulse with fluence 
of the order of $\sim 70 \mu$J~cm$^{-2}$ and photon energy above the gap.
A detailed discussion of the initial conditions in reported in the SI. 
}
The lattice is initially at thermal equilibrium with phonon populations $n_{\bq
\nu}$ determined from  Bose-Einstein statistics via $n_{\bq \nu} =
[e^{\hbar\omega_{\bq\nu} / k_B T_0}-1]^{-1}$ with $T_0 =$ 300 K. 

For these initial conditions, we investigate the coupled dynamics of electrons
and phonons based on the solution of the time-dependent Boltzmann equation
(TDBE):\cite{TDBE2}
\begin{align}
\partial_t f_{n\bk}(t) &= \Gamma^{\rm ep}[f_{n\bk}(t),n_{\bq \nu}(t)] \label{TDBE1} \\
\partial_t n_{\bq \nu}(t) &= \Gamma^{\rm pe} [f_{n\bk}(t),n_{\bq \nu}(t)] + \Gamma^{\rm pp}[n_{\bq \nu}(t)] \label{TDBE2}
\end{align} 
with $\partial_t = \partial / \partial t$. $\Gamma^{\rm ep}$,
$\Gamma^{\rm pe}$, and $\Gamma^{\rm pp}$ are the collision integrals due to
electron-phonon (e-ph), phonon-electron (ph-e) and phonon-phonon (ph-ph)
interactions, respectively, which are evaluated entirely from first principles.
Explicit expressions are reported in the Supporting Information. Radiative
recombination and electron-hole interactions are neglected hereafter. 
{\color{black} 
At room temperature the scattering due to electron-electron interaction 
is negligible for carriers in the vicinity of the band edges\cite{WSe2_NEGF} 
and it is therefore omitted in our simulations (see SI for a detailed discussion of this approximation).}
In the TDBE the ultrafast dynamics of phonons, electrons, and holes is treated on an
equal footing. In particular, momentum, time, and band resolutions are
retained, providing a more detailed picture of the system dynamics as compared
to effective temperature models,\cite{nonthermal_MTM,nonthermal_MTM2} and the
relaxation-time approximation.\cite{RTA} We implemented Eqs.~\eqref{TDBE1} and
\eqref{TDBE2} and the collision integrals  $\Gamma^{\rm ep}$, $\Gamma^{\rm
pe}$, and $\Gamma^{\rm pp}$  in the {\tt EPW} code.\cite{EPW} The time
derivative is discretized based on Runge-Kutta method. We used a time step of
1~fs for a total simulation time of 6~ps. All computational details are
reported in the SI. 
 
    We proceed to examine the valley depolarization dynamics of electrons and
holes.  The electron distribution function $f_{n{\bf k}}(t)$ for the conduction
band -- reported in Fig.~\ref{electron_dynamics}~(c) and (d) for $t=300$ and
900~fs, respectively -- reveals that valley polarization decays within 1~ps and
an isotropic occupation of the K and K$'$ valleys is re-established by
phonon-assisted intervalley electron scattering from K to K$'$.  Conversely,
the depolarization dynamics of the valence band occurs on longer time scales,
as reflected by the corresponding distribution function $f_{n{\bf k}}$ reported
in Fig.~\ref{electron_dynamics}~(f) and (g) for $t = 300$ and 900 fs. This
behaviour can be understood by noticing that phonon-assisted transitions from K
to K$'$ in the valence band involve either a spin-flip process
(K$\rightarrow$K$'$) or an intermediate transition to the $\Gamma$ valley
(K$\rightarrow\Gamma\rightarrow$K$'$). These processes occur at a slower rate,
as compared to direct K$\rightarrow$K$'$ transitions in the conduction band,
which do not require spin reversal. To further illustrate the timescales of the
valley depolarization, we report in  Fig.~\ref{electron_dynamics}~(h) and (i)
the electron distribution function $f_{n{\bf k}}(t)$ for momenta in the
vicinity of $\krm$ and $\kprime$ in the valence and conduction band,
respectively. {\color{black} An exponential fit to the curves yields decaying
times 2 ps (150 fs) for the valence (conduction) band. %These values are in
excellent agreement with earlier tr-ARPES
experiments,\cite{ws2_valley_lifetime} and first-principles calculations based
on the relaxation time approximation.\cite{MX2_eph_soc} These values are
consistent with previous first-principles calculations based on relaxation time
approximation.\cite{MX2_eph_soc} Moreover, the decaying time for the conduction
band is in excellent agreement with earlier tr-ARPES
experiments.\cite{ws2_valley_lifetime}}
    
\begin{figure*}[t]
\begin{center}
\includegraphics[width=0.98\textwidth]{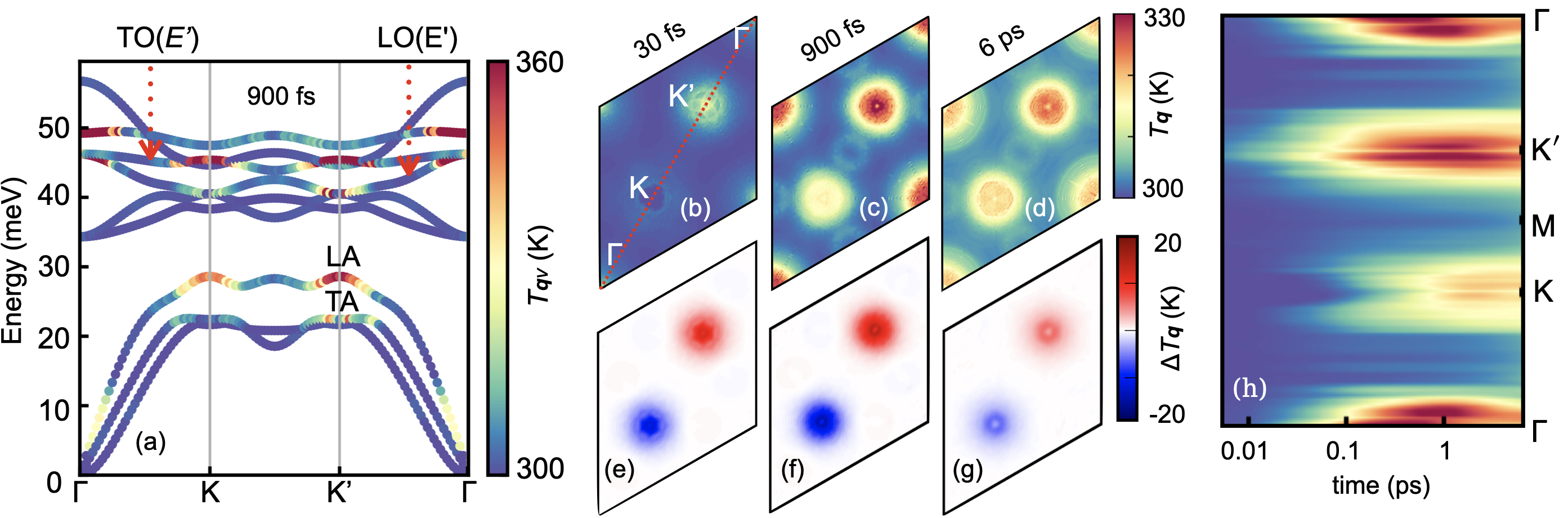}
\end{center}
\caption{\label{phonon_dynamics} Nonequilibrium dynamics of valley-polarized
phonons (a) Mode- and momentum-resolved effective phonon temperature at
$t=900$~fs superimposed as color coding onto the phonon dispersion. (b)-(d)
Momentum-resolved effective vibrational temperature $\tilde{T}_{\bq}$ for
crystal momenta within the Brillouin zone  at $t= 30$, 900, and 6000 fs,
respectively. (e)-(g) Momentum-resolved differential vibrational temperature
$\Delta \widetilde{T}_{\bq} =
\widetilde{T}_{\bq}^{\circlearrowleft}-\widetilde{T}_{\bq}^{\circlearrowright}$
upon switching the light pump from left- to right-hand polarized for the same
times as in panels (b)-(d). (h) Time-dependent average phonon temperature
$\tilde{T}_{\bq}$ for modes along the high-symmetry path
$\Gamma$-K-M-K$'$-$\Gamma$ in the Brillouin zone, marked by a dashed line in
panel (b).}
\end{figure*}
	
The valley depolarization dynamics discussed above has important implications
for the nonequilibrium phonon distributions, which we analyze in detail
hereafter.  To inspect the non-equilibrium phonon population established upon
thermalization of valley-polarized electrons and holes, we define the
effective temperature:\cite{MoS2_TDBE}
\begin{equation}
T_{\bq \nu} =\hbar \omega_{\bq\nu} [{k_{\rm B}{\rm ln}(1+n_{\bq \nu}^{-1})}]^{-1}\quad.
\end{equation}
where $k_{\rm B}$ is the Boltzmann constant, $\omega_{\bq\nu}$ the phonon
frequency, and $n_{\bq \nu}$ is obtained from Eq.~\eqref{TDBE2}.  All phonons
exhibit identical effective temperatures at thermal equilibrium ($T_{\bq
\nu}=$~constant), whereas $T_{\bq \nu}$ may change for different vibrational
modes out of equilibrium. To exemplify the non-equilibrium vibrational states
of the lattice induced by the electron-phonon interaction, we report in
Fig.~\ref{phonon_dynamics}~(a) the effective temperature $T_{\bq \nu}$ at
$t=900$~fs, superimposed as a color coding to the phonon dispersion. The
enhanced temperature around $\Gamma$, K, and K$'$ for the longitudinal acoustic
(LA) and transverse optical (TO) phonons reflects the enhanced bosonic
occupation of these modes, indicating that they are emitted at a higher rate
during the thermalization dynamics of photoexcited carriers. See the SI for the
phonon labelling convention.  The mode-resolved effective temperature $T_{\bq
\nu}$ at 900 fs in Fig.~\ref{phonon_dynamics}(a) indicates that electron-phonon
interactions for photo-excited carriers are highly mode selective.  In
particular, intravalley scattering is governed by the emission of
long-wavelength ($\bq\simeq0$) ZO($A_1 '$), LO($E'$) and LA phonons,
consistently with earlier theoretical and experimental
studies.\cite{MX2_eph,Jin2014,MoS2_symmetry,Li2013,Molina-Sanchez2016,MoSe2_LA(K)}
Conversely, intervalley scattering and the ensuing valley depolarization is
dominated by emission of LA, TA, LO($E'$) and TO($E'$) phonons. 

   To visualize the nonequilibrium phonon populations with full momentum
resolution across the Brillouin zone, we report in
Figs.~\ref{phonon_dynamics}~(b-d)  the mode-averaged effective phonon
temperature $\widetilde{T}_{\bq} = N_{\rm ph}^{-1}\sum_{\nu} T_{\bq \nu}$,
{\color{black} where $N_{\rm ph}$ is number of distinct vibrational modes.}
Already at $t=$30~fs [Figs.~\ref{phonon_dynamics}~(b)] the lattice enters a
nonequilibrium regime characterized by an anisotropy in
the phonon temperature at the K and K$'$ valleys.  The phonon valley anisotropy
reaches a maximum within 1 ps and persists for several picoseconds, as
illustrated in Fig.~\ref{phonon_dynamics}~(h), where the phonon temperature
$\tilde T_{\bq}$ along the $\Gamma$-K-K$'$-$\Gamma$ path is reported as a
function of time. On longer timescales, the anisotropic population of K and
K$'$ phonons is gradually suppressed due to phonon-phonon scattering.
\cite{dfsct_WSe2} This behaviour stems from phase-space constraints in the
phonon emissions throughout the depolarization dynamics of conduction and
valence bands.  In particular, for electron scattering in the conduction band
($\krm \rightarrow \kprime$), phonon emission processes are constrained to the
vicinity of the $\kprime$ high-symmetry point due to momentum conservation.
Hole scattering in the valence band (K$\rightarrow\Gamma\rightarrow$K$'$) is
also dominated by phonon emissions at K$'$, whereas transitions along ($\krm
\rightarrow \kprime$) are low-rate spin-flip processes.  These findings provide
strong indication that a valley polarization in the phonon population can be
established in monolayer MoS$_2$ (and isostructural compounds) following
photoexcitation with circularly polarized light, and it persists for longer
timescales than the electron valley polarization.
    
   Below we show that light helicity enables to selectively reverse the
population of valley-polarized phonons.  We report in
Figs.~\ref{phonon_dynamics}~(e-g) the dichroic vibrational temperature defined
as $\Delta \widetilde{T}_{\bq} =
\widetilde{T}_{\bq}^{\circlearrowleft}-\widetilde{T}_{\bq}^{\circlearrowright}$,
for $t=30$, 900, and 6000~fs. Here, $\widetilde{T}_{\bq}^{\circlearrowleft}$ ($
\widetilde{T}_{\bq}^{\circlearrowright}$) denotes the vibrational temperature
established by considering an initial population of electrons and holes at the
K (K$'$) valley as a result of photoexcitation with left-handed  (right-handed)
circularly-polarized light.  Figures~\ref{phonon_dynamics}~(e-g) reveal that
the valley-polarized phonon population at K and K$'$ is fully reversed by
considering opposite circular polarization for the driving pulse. The helicity
of circularly-polarized light pulses thus enables to selectively excite phonons
at either the K or K$'$ high-symmetry points, leading to the emergence of
circular dichroism in the vibrational dynamics of monolayer MoS$_2$. 
   
   The nonequilibrium vibrational states illustrated in
Fig.~\ref{phonon_dynamics} are characterized by nontrivial chirality, which can
be quantified by introducing the total phonon circular polarization $L =
N_p^{-1}\sum_{\bq\nu} n_{\bq\nu} L_{\bq\nu}^{z}$. Here $L_{\bq\nu}^{z}$ denotes
the circular polarization of a single phonon mode \cite{chiral_phonon_theory}
obtained by projecting the phonon eigenvector on a circularly polarized basis
(see SI).  $L_{\bq\nu}^{z}>0$ ($L_{\bq\nu}^{z}<0$) corresponds to chiral
phonons with right-(left-)handed circularly polarization, and $L_{\bq\nu}^{z}=
0$ to nonchiral phonons. The strongly-coupled LO($E'$) and TA modes at K$'$ are
strongly chiral with $L_{\bq\nu}^{z} = 1$ and $-1$, respectively, whereas the
TO($E'$) and LA modes have trivial chirality ($L_{\bq\nu}^{z} \simeq 0$).
Time-reversal symmetry requires a total vanishing chirality ($L=0$) at thermal
equilibrium, whereas $L\neq 0$ can only arise if time-reversal symmetry is
lifted.  As illustrated in Fig.~S3 (b), the anisotropic phonon population at K
and K$'$ that accompanies the emergence of valley-polarized phonons breaks
time-reversal symmetry, and it confers to the nonequilibrium dynamics of the
lattice a nontrivial chiral character ($L\neq 0$) which persists until
thermalization of the K and K$'$ valley phonons.
  
\begin{figure*}[t]
\begin{center}
\includegraphics[width=0.98\textwidth]{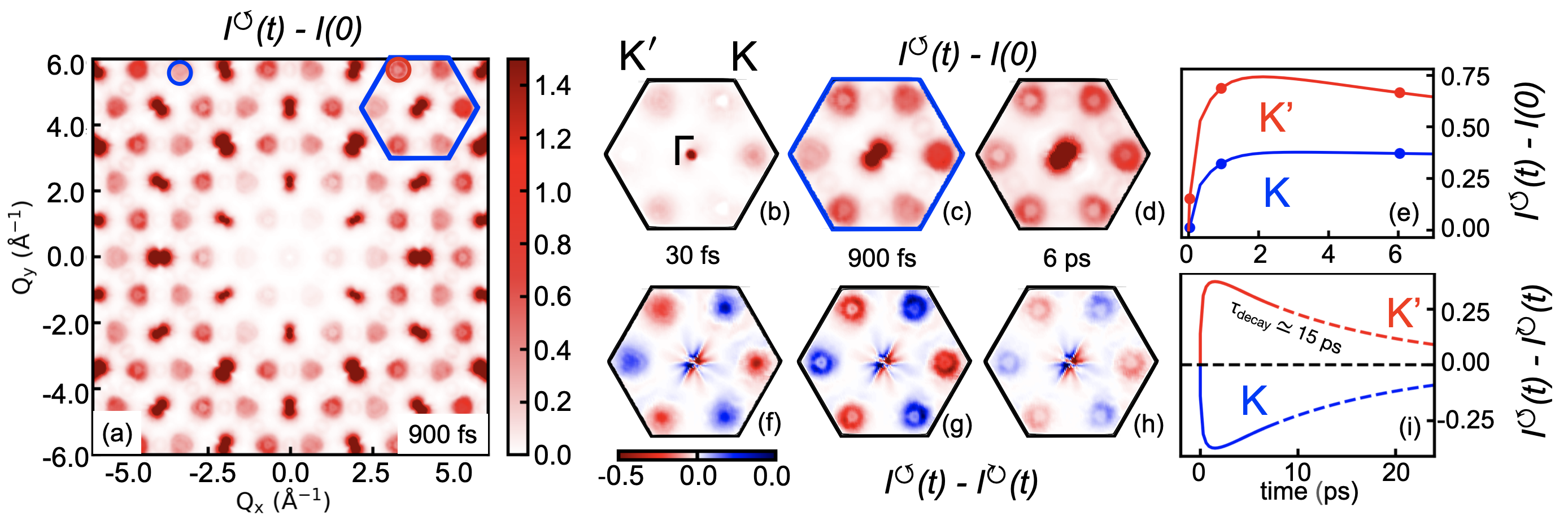}
\end{center}
\caption{\label{diffuse_scattering} Vibrational dichorism of valley-polarized
phonons. (a) Differential diffuse scattering intensity $\Delta I(t) =
I^{\circlearrowleft}(t)-I(t=0)$ for $t=900$~fs after photoexcitation. (b)-(d)
$\Delta I(t)$ for momenta in the hexagonal region marked in panel (a) and $t=
30$, 900, 6000 fs. (e) Time dependence of $\Delta I(t)$ for momenta around K
and K$'$. (f)-(h) Dichroic diffuse scattering intensity
$D(t)=I^{\circlearrowleft}(t)-I^{\circlearrowright}(t)$ for $t=30$, 900, 6000
fs. (i) Time dependence of  $D(t)$ in the vicinity of K and $\kprime$. The
dashed line is an exponential fit. }
\end{figure*}

    Ultrafast electron diffuse scattering (UEDS) enables to directly image the
nonequilibrium phonon populations with time and momentum
resolutions.\cite{UEDS_RMP,graphene_UEDS,dfsct_WSe2,dfsct_disca1,dfsct_disca2}
UEDS has been widely applied in the past to layered and 2D
materials,\cite{dfsct_WSe2,bP_UEDS,MoS2_UDES} making it suitable to explore the
emergence of chiral phonons in real time in TMDs.\cite{britt2023ultrafast}  We
thus expect it to provide a suitable route to experimentally detect occurrences
of vibrational dichroism. In the following, we show that the nonequilibrium
phonon populations established by absorption of circularly-polarized light give
rise to distinctive fingerprints in the measured UEDS signal, characterized by
a reversal of UEDS intensity upon switching of light polarization. The UEDS
intensity is directly related to the dynamical structure factor, which can be
expressed as:
\begin{equation}\label{eq:I}
I(\bQ,t) \propto \sum_{\nu} |F_{1\nu}(\bQ) |^2 \frac{n_{\bq \nu}(t)+1/2}{\hbar\omega_{\bq \nu}} \quad .
\end{equation}
    Here, we retain only the one-phonon contribution, $F_{1\nu}(\bQ) $ is the
one-phonon structure factor {\color{black} (see SI)}, and $\bQ = \bq + \bG $ is
the transferred momentum, with $\bq$ and $\bG$ denoting the phonon momentum and
the
reciprocal lattice vector, respectively.  To unveil the fingerprints of
vibrational dichroism in UEDS, we evaluate Eq.~\eqref{eq:I} from first
principles using the time-dependent phonon occupation $n_{\bq \nu}$ obtained
from Eq.~\eqref{TDBE2}. The differential diffuse scattering intensity $\Delta
I(\bQ,t) = I^{\circlearrowleft}(\bQ,t)-I(\bQ,t=0)$ is reported in
Fig.~\ref{diffuse_scattering}~(a) for $t=900$~fs, and  $\circlearrowleft$
denotes photoexcitation with left-handed circularly polarized light. Positive
values of the color scale reflect an overall enhancement of the UEDS signal
which results from the increased phonon population. 
    
    We concentrate in the following on the crystal momenta within the blue
hexagon in Fig.~\ref{diffuse_scattering}~(a), which marks the Brillouin zone
corresponding to the $210$ reciprocal lattice vector.
Figures~\ref{diffuse_scattering}~(b-d) depict $\Delta I(\bQ,t)$ for $t=30$,
900, and 6000~fs, respectively, for transferred momenta in the hexagon.
Changes of $\Delta I(\bQ,t)$ at the center of the hexagon stem from changes of
the phonon population at $\Gamma$, whereas the intensity at the zone edges
arise from the excitation of K and K$'$ phonons as marked in
Fig.~\ref{diffuse_scattering}~(b).  Momenta corresponding to the K$'$ (K)
high-symmetry point are characterized by larger (lower) diffuse scattering
intensity for $t=$30 and 900~fs, thus providing evidence of a strong dichroic
vibrational response. On longer timescales, $\Delta I$ decreases due to the
onset of phonon-phonon scattering that drives the lattice towards thermal
equilibrium. However, clear signatures of dichroism persist beyond 5~ps.  The
full time-dependence of the differential diffuse scattering intensity is
illustrated in Fig.~\ref{diffuse_scattering}~(e), where we report  $\Delta I$
for momenta in the vicinity of the K and K$'$ valleys. The blue (red) line has
been obtained by averaging the differential UEDS intensity $\Delta I$ over
momenta within the blue  (red) circle in panel
Fig.~\ref{diffuse_scattering}~(a), which corresponds to the K (K$'$) point in
the Brillouin zone. These high-symmetry points are related by time-reversal
symmetry at thermal equilibrium and they exhibit identical UEDS intensities
before pump [$\Delta I ({\rm K},0) = \Delta I ({\rm K}',0) $]. After pump,
conversely, time-reversal symmetry is broken by the anisotropic population of K
and K$'$ phonons, leading to different UEDS signals at K and K$'$. 

To further illustrate the fingerprints of vibrational dichroism in UEDS, we
report in Fig.~\ref{diffuse_scattering}~(f-h) the dichroic diffuse scattering
intensity defined as
$D(\bQ,t)=I^{\circlearrowleft}(\bQ,t)-I^{\circlearrowright}(\bQ,t)$ for $t=30$,
900, and 6000~fs. Red (blue) colors mark the increase (decrease) of UEDS
intensity upon reversal of circular light polarization.  The time-dependence of
$D(\bQ,t)$ is illustrated in Fig.~\ref{diffuse_scattering}~(i) for momenta in
the K and K$'$ valleys.  By an exponential fit, we extract a decaying time
$\tau_{\rm decay}=15$~ps from the dichroic signal $D(\bQ,t)$, which quantifies
the characteristic timescales on which the polarization of valley phonons
persists. 
    
In conclusion, we reported an ab-initio investigation of the nonequilibrium
lattice dynamics of monolayer MoS$_2$ resulting from the thermalization of a
valley-polarized electronic excitation.  Our study reveals the emergence of
valley-polarized phonon populations in the nonequilibrium phonon dynamics
following absorption of circularly-polarized photons, which are localized
either at the K or K$'$ high-symmetry points depending on light helicity.
Simulations of the dynamical structure factor further reveal distinctive
fingerprints of vibrational dichroism in ultrafast diffuse scattering
experiments that can persist over timescales beyond 10~ps, offering an
accessible opportunity for experimental detection of this phenomenon.
 
Overall, these results unveil a novel route to achieve control of the phonon
dynamics using circular light polarization. In particular, the dichroic lattice
response may offer an unexplored opportunity to directly drive
the excitation of chiral phonons, thereby lifting time-reversal symmetry, with
possible implications for the topological properties of the lattice.  The
dichroism of valley phonons emerges as a universal feature of the
nonequilibrium lattice dynamics, which we also expect in other
noncentrosymmetric hexagonal crystals as, e.g., other TMDs (e.g., WS$_2$ and
MoSe$_2$ monolayer), hexagonal BN, and the novel family of 2D
metals.\cite{Briggs2020}
 
{\bf Supporting Information}\\ 
The Supporting Information is available free of charge at {\tt
https://link.provided.by.publisher}.

Computational details; Initial conditions of the ultrafast dynamics
simulations; expressions for the collision integrals; analysis of
phonon chirality and selection rules; time-, momentum-, and mode-resoveld
phonon temperatures (PDF).
	
{\bf Acknowledgments.}  
This project has been funded by the Deutsche Forschungsgemeinschaft (DFG) --
project numbers 443988403. We acknowledge discussions with Marios Zacharias,
Tristan L. Britt, Bradley J.  Siwick, and Sivan Refaely-Abramson. 

\bibliography{references}
\end{document}